\documentstyle[multicol,prl,aps,psfig,floats]{revtex}
\begin{document}
\tightenlines
\title{Comment on ``Absence of electron dephasing at zero temperature''}
\author{Dmitri S. Golubev$^{1,3}$, Andrei D. Zaikin$^{2,3}$ and 
Gerd Sch\"on$^{1,2}$}
\address{$^1$Institut f\"ur Theoretische Festk\"orperphysik,
Universit\"at Karlsruhe, 76128 Karlsruhe, Germany \\
$^2$Forschungszentrum Karlsruhe, Institut f\"ur Nanotechnologie,
76021 Karlsruhe, Germany\\
$^3$I.E.Tamm Department of Theoretical Physics, P.N.Lebedev
Physics Institute, Leninskii pr. 53, 119991 Moscow, Russia}

\maketitle

\begin{abstract}

The recent claim by Kirkpatrick and Belitz
\cite{KB} that Ward identities could be used to prove
the absence of electron dephasing at $T=0$ contains serious flaws.
These authors try to draw conclusions about dephasing
from an analysis of the diffuson, which is not sensitive to this
process. The Cooperon, which does contain this information, 
is analyzed only in time reversal symmetric situations, which {\it by
assumption} excludes any relaxation and dephasing. Hence,
their analysis remains inconclusive for the problem in question. 

\end{abstract}


\begin{multicols}{2}
 

The prediction of quantum dephasing due to electron-electron 
interactions at low temperature continues to attract attention. In 
response to our work\cite{GZ} some authors have been searching for 
errors and/or invalid approximations in our analysis. We studied 
many of these -- published or unpublished -- claims and could discard 
them as being in error and/or inapplicable\cite{GZ,GZS}. This applies 
also for the recently published critique by Aleiner et al.\cite{AAV}, 
to which we will respond in a separate publication.

Other authors try to prove the desired result, vanishing dephasing at 
$T=0$, by general arguments. 
Recently Kirkpatrick and Belitz\cite{KB} (KB) argued that because of  
Ward identities for electrons in disordered conductors ``neither a 
Coulomb interaction nor a short-ranged
model interaction can lead to phase breaking at zero temperature
in spatial dimensions $d>2$''. Here we show that 
KB's analysis is incorrect and inconclusive in several points:

{\bf KB do not distinguish adequately between Cooperon and diffuson.}
The bulk of their paper\cite{KB} is devoted to an elaborate
derivation of the well known
fact, that the diffuson does not decay (i.e.~is massless) at $T=0$. 
[Actually, as we will discuss below, also this part of KB's analysis 
contains serious drawbacks and, hence, cannot be accepted. However,
for the moment we proceed with the observation that
the diffuson is massless at $T=0$.] 
The crucial question then remains whether this property of the
diffuson has any implications for electron decoherence. The
point is that the {\it Cooperon}, rather than the diffuson, contains the 
information about the electron dephasing time $\tau_{\varphi}$. KB do
not evaluate the Cooperon other than in ``the presence of time 
reversal invariance''. In this case the time dependences of the
diffuson and the Cooperon are the same, but at the same time all
relaxation and dephasing processes are excluded {\it by assumption}.  This
step constitutes KB's  major error.  Extending this part of KB's 
arguments one could ``prove'' that no irreversible phenomena, for
instance dissipation and, of course, dephasing exist in
quantum mechanics.

KB appear not to appreciate the fact that time reversal invariance 
is broken for an electron interacting with a bath, in the 
present problem with the bath produced by all other
electrons\cite{FN}. 
This is true at any temperature, including $T=0$, and can be observed already
at the level of {\it exact} equations of motion\cite{FN2}. When the time 
reversal symmetry is broken the dynamics of diffusons and Cooperons is 
fundamentally different. No conclusions about dephasing can be
derived from the former.

{\bf KB consider primarily short-range interactions.}
In eq. (3.9b) of \cite{KB} we observe that KB analyze a
Thomas-Fermi screened interaction rather than the {\it unscreened} 
Coulomb interaction\cite{GZ}.
However, the latter model\cite{GZ} is generally
accepted\cite{Aronov,Schmid} as physically 
meaningful for the problem in question. The two models can lead to
qualitatively differing conclusions.

{\bf For long-range interactions no meaningful results follow from 
the analysis of KB.}
One might expect that the consequences of long-range 
interactions could be recovered from the work of KB\cite{KB} if one sets
the inverse screening length $\kappa$ in eq. (3.9b) to zero. However,  
following their derivation to eq. (3.20) one observes that the term 
$W^{(dc)}$ ``vanishes continuously as $\kappa \to 0$ for $d>2$''. 
Combining this result with eq. (3.19a) one finds that for $d>2$ the 
diffuson is not affected by the interaction and does not decay at any 
temperature. Due to KB's assumption (3.26), this applies for the
Cooperon as well. Thus, for  $\kappa \to 0$ the analysis of
Ref.~\cite{KB} predicts the absence of electron dephasing at any
temperature. 

A more accurate procedure (not performed by KB) amounts to 
using the Fourier transform of (3.9b) $v_{sc}(q)=4\pi
e^2/(q^2+\kappa^2)$ (in $d=3$) rather than their eq.~(3.9c). In the latter
form KB expressed the strength of the interaction ($\propto e^2$) 
via $\kappa^2 =4\pi e^2N_F$ in terms of the screening length. As a
result the limit $\kappa \to 0$, corresponding to long range
interaction, simultaneously implies vanishing strength of the
interaction. It is therefore 
not surprising that the diffuson is unaffected by interactions.
When proceeding with the proper form of the interaction to KB's eq.~(3.20) 
one arrives for $\kappa \to 0$ at a meaningless divergence $W^{(dc)} 
\propto 1/\kappa^{3}$. Thus, for unscreened Coulomb interaction the 
analysis \cite{KB} fails, and no conclusion  can be
drawn from it even about the diffuson.

On the other hand, the property that the diffuson does not decay at 
$T=0$ {\it does} follow (trivially) from our formalism \cite{GZ,GZS}. 
Hence, KB's claim that our ``calculations and arguments violate an 
exact Ward identity'' and therefore are  ``not even internally
consistent'' lacks any substantiation and is in error.

\end{multicols}


\begin{references}
\bibitem{FN2} See, e.g., eqs. (24), (25) in the first Ref. \cite{GZ}
or eq. (20) in ref. \cite{GZS} where the effect of the electronic bath
is expressed in terms of the fluctuating fields $V_{1,2}$.
\end{references}
\end{document}